\documentclass[sigconf]{acmart}

\AtBeginDocument{%
  }

\usepackage{xcolor}
\usepackage{pifont}
\usepackage{booktabs}
\usepackage[table]{xcolor}
\usepackage{listings}
\usepackage{etoolbox}
\usepackage{multirow}
\usepackage[table]{xcolor} 
\usepackage{xspace}
\newcommand{\hltm}{\textsc{HLTM}\xspace}

\usepackage{makecell}
\usepackage{balance}

\copyrightyear{2026}
\acmYear{2026}
\setcopyright{cc}
\setcctype{by-nc-nd}
\acmConference[KDD 2026] {Proceedings of the 32nd ACM SIGKDD Conference on Knowledge Discovery and Data Mining V.2}{August 9--13, 2026}{Jeju Island, Republic of Korea.}
\acmBooktitle{Proceedings of the 32nd ACM SIGKDD Conference on Knowledge Discovery and Data Mining V.2 (KDD 2026), August 9--13, 2026, Jeju Island, Republic of Korea}
\acmISBN{979-8-4007-2259-2/2026/08}
\acmDOI{10.1145/3770855.3818432}
\begin{document}

\title{Hierarchical Long-Term Semantic Memory for LinkedIn's Hiring Agent}


\author{Zhentao Xu}
\email{zhexu@linkedin.com}
\orcid{0009-0000-5827-7879}
\affiliation{
  \institution{LinkedIn Corporation}
  \city{Sunnyvale}
  \state{CA}
  \country{USA}
}

\author{Shangjin Zhang}
\email{shanzhang@linkedin.com}
\orcid{0009-0006-9318-0363}
\affiliation{
  \institution{LinkedIn Corporation}
  \city{Sunnyvale}
  \state{CA}
  \country{USA}
}

\author{Emirhan Poyraz}
\email{epoyraz@linkedin.com}
\orcid{0000-0002-1863-7419}
\affiliation{
  \institution{LinkedIn Corporation}
  \city{New York}
  \state{NY}
  \country{USA}
}

\author{Yvonne Li}
\email{yiwli@linkedin.com}
\orcid{0009-0005-8610-6168}
\affiliation{
  \institution{LinkedIn Corporation}
  \city{Sunnyvale}
  \state{CA}
  \country{USA}
}

\author{Ye Jin}
\email{yejin@linkedin.com}
\orcid{0000-0003-3882-952X}
\affiliation{
  \institution{LinkedIn Corporation}
  \city{Sunnyvale}
  \state{CA}
  \country{USA}
}

\author{Xie Lu}
\email{xilu@linkedin.com}
\orcid{0009-0005-9190-915X}
\affiliation{
  \institution{LinkedIn Corporation}
  \city{Sunnyvale}
  \state{CA}
  \country{USA}
}

\author{Xiaoyang Gu}
\email{xgu@linkedin.com}
\orcid{0009-0004-7882-4103}
\affiliation{
  \institution{LinkedIn Corporation}
  \city{Sunnyvale}
  \state{CA}
  \country{USA}
}

\author{Karthik Ramgopal}
\email{kramgopa@linkedin.com}
\orcid{0009-0006-0647-2385}
\affiliation{
  \institution{LinkedIn Corporation}
  \city{Sunnyvale}
  \state{CA}
  \country{USA}
}

\author{Praveen Kumar Bodigutla}
\email{pbodigutla@linkedin.com}
\orcid{0000-0003-3244-467X}
\affiliation{
  \institution{LinkedIn Corporation}
  \city{Sunnyvale}
  \state{CA}
  \country{USA}
}

\author{Xiaofeng Wang}
\email{xiaofwang@linkedin.com}
\orcid{0009-0009-5648-4183}
\affiliation{
  \institution{LinkedIn Corporation}
  \city{New York}
  \state{NY}
  \country{USA}
}

\renewcommand{\shortauthors}{Zhentao Xu et al.}

\begin{abstract}
    Large Language Model (LLM) agents are increasingly used in real-world products, where personalized and context-aware user interactions are essential. A central enabler of such capabilities is the agent's long-term semantic memory system, which extracts implicit and explicit signals from noisy longitudinal behavioral data, stores them in a structured form, and supports low-latency retrieval. Building industrial-grade long-term memory for LLM agents raises five challenges: scalability, low-latency retrieval, privacy constraints, adaptability, and observability. We introduce the \textit{Hierarchical Long-Term Semantic Memory} (\hltm) framework, which organizes textual data into a schema-aligned memory tree that captures semantic knowledge at multiple levels of granularity, enabling scalable ingestion, privacy-aware storage, low-latency retrieval, and transparent provenance; \hltm further incorporates an adaptation mechanism to generalize across diverse use cases. Extensive evaluations on LinkedIn's Hiring Assistant show that \hltm improves answer correctness by more than 5\% and retrieval F1 by more than 10\%, while significantly advancing the Pareto frontier between query and indexing latency. \hltm has been fully deployed in LinkedIn's Hiring Assistant to power core personalization features in production hiring workflows.
\end{abstract}




\maketitle

\section{Introduction}
Large language models (LLMs) such as GPT~\cite{achiam2023gpt}, Claude~\cite{anthropic_homepage}, and Gemini~\cite{comanici2025gemini} have demonstrated strong performance across a wide range of tasks in real-world agent systems, including question answering, code generation, and retrieval-augmented search~\cite{kamalloo2023evaluating,zan2023large,zhu2025large}. Modern LLM-based agents~\cite{wang2024survey} augment these models with external tools~\cite{schick2023toolformer} and carefully designed workflows to enhance their reasoning and planning capabilities~\cite{wei2022chain,yao2023tree}. For example, LinkedIn’s Hiring Assistant~\cite{gu2025hiringassistant} is an AI recruiting agent that helps recruiters source, evaluate, and outreach to candidates at scale. Agentic memory systems~\cite{zhang2025survey,zhong2024memorybank,zhang2025g,xu2025mem,jimenez2024hipporag,chhikara2025mem0} support an agent's long-term interaction with its environment by persisting user-specific signals from historical data and leveraging them to personalize future agent--user exchanges. For instance, LinkedIn’s Hiring Assistant~\cite{gu2025hiringassistant} needs to extract personalized hiring preferences from recruiters' past projects and activities, and recall this recruiter-specific memory to generate tailored responses in future interactions. Achieving this personalization at scale requires a memory system that reliably extracts implicit and explicit signals from noisy longitudinal data, stores them in structured form, and retrieves them efficiently during real-time conversations.

Recent work has advanced agents’ long-context capabilities by introducing explicit memory modules. Systems such as MemGPT, Mem0, MemoryBank, ReadAgent, MemOS, SimpleMem, and SCM~\cite{packer2023memgpt,chhikara2025mem0,zhong2024memorybank,lee2024human,li2025memos,liu2026simplemem,wang2023scm} extend LLMs with external stores that swap, persist, and selectively retrieve interaction traces beyond the native context window. Complementary efforts explore more structured representations: GraphRAG, HippoRAG, A-Mem, and G-Memory~\cite{edge2024local,jimenez2024hipporag,xu2025mem,zhang2025g} organize knowledge into graphs to support multi-hop reasoning, while RAPTOR, TreeRAG, and MemTree~\cite{sarthi2024raptor,tao2025treerag,rezazadeh2024isolated} exploit hierarchical structure for scalable long-context retrieval. Despite these advances, designing long-term semantic memory for large-scale, industrial-grade agents remains challenging across several key dimensions:
\begin{itemize}
\item \textbf{Challenge 1: Scalability} A production memory system must ingest and process large-scale structured or unstructured data from diverse sources and formats. This demands compact memory representations, highly parallelizable ingestion pipelines, and lossless incremental updates.
\item \textbf{Challenge 2: Low-latency Retrieval} Many scenarios impose strict latency constraints. For example, chat-based agents must respond within tight budgets to maintain a fluent conversation, leaving little room for expensive retrieval or oversized context windows at serving time. As a result, most computation is ideally amortized offline or nearline, with only a lean, predictable retrieval path online.
\item \textbf{Challenge 3: Privacy Constraints} Enterprise agents must comply with strict data-governance policies (e.g., GDPR~\cite{gdpreu_guide}). With these policies, the memory must enforce strict isolation, preventing cross-tenant retrieval, and support clean deletion when required.
\item \textbf{Challenge 4: Adaptability} The memory framework must adapt continuously to shifting query patterns, rather than depending on brittle hand-crafted rules or feature-specific engineering pipelines.
\item \textbf{Challenge 5: Observability} The memory system should preserve references to its underlying information sources and surface them during retrieval, enabling inspection, debugging, and governance. It should also remain consistent as the source of truth (SOT) evolves (e.g., edits, backfills, or deletions), preventing stale or contradictory memories.

\end{itemize}

\begin{figure}[t]
    \centering
    \includegraphics[width=1\linewidth]{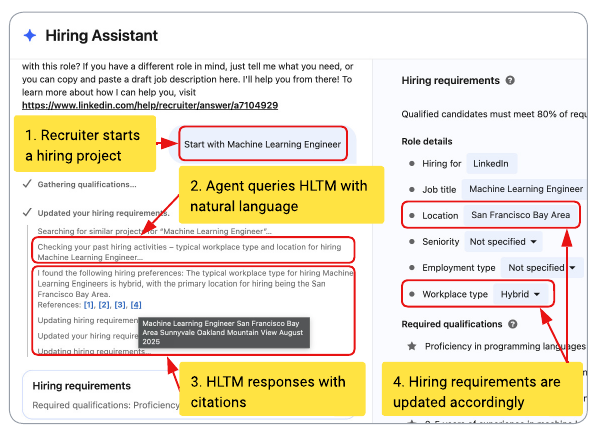}
    \caption{LinkedIn's Hiring Assistant UI with \hltm: a recruiter initiates a hiring project; the Hiring Assistant queries \hltm in natural language to retrieve preference signals, then uses the returned information to update structured hiring requirements.}
    \Description{Screenshot of LinkedIn's Hiring Assistant user interface, showing a recruiter starting a hiring project. The assistant issues a natural-language query to the \hltm memory system, receives preference signals, and uses them to populate structured hiring requirement fields.}
    \label{fig:LiHA_UI}
\end{figure}

\begin{figure}[t]
    \centering
    \includegraphics[width=1\linewidth]{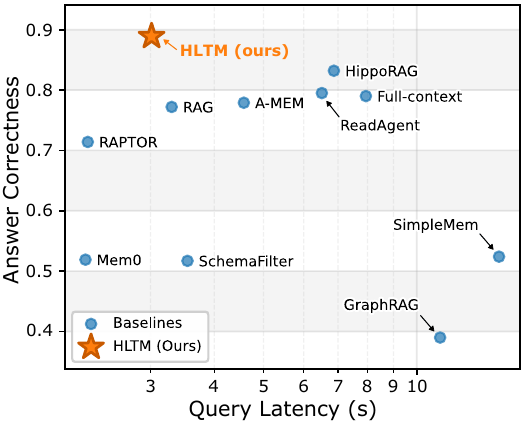}
    \caption{Performance–latency trade-off across evaluated methods on LinkedIn’s Hiring Assistant dataset. The \hltm consistently achieves higher answer correctness while maintaining lower latency compared to baseline approaches.}
    \Description{Scatter plot of answer correctness versus query latency for \hltm and baseline methods on LinkedIn's Hiring Assistant dataset. \hltm appears in the upper-left region, achieving the highest correctness at a low latency of about three seconds, while other baselines either have lower correctness or substantially higher latency.}
    \label{fig:performance-vs-latency}
\end{figure}

To address these challenges, we propose \hltm, a hierarchical long-term semantic memory framework for industrial LLM agents. \hltm organizes memory into a tree-structured index that supports massively parallel execution and lossless incremental ingestion, generates multi-granularity memory representations for low-latency retrieval at serving time, and supports enhanced privacy policies by aligning the memory hierarchy with business scopes such as seats and projects. The framework is adaptable through automatic query pattern analysis and knowledge extraction. Each memory node is bound to a stable business identifier, providing explicit provenance and observability for inspection, debugging, and governance. \hltm has been fully deployed in production as the long-term semantic memory of LinkedIn's Hiring Assistant~\cite{gu2025hiringassistant} (Figure~\ref{fig:LiHA_UI}). Over six months of real-world operation, it has demonstrated robust, reliable, enterprise-scale performance at controlled latency and cost (Figure~\ref{fig:performance-vs-latency}).

Our main contributions are summarized below:
\begin{itemize}
    \item \textbf{Production-scale long-term memory for enterprise agents:} We present an industrial-scale memory architecture that scales to millions of documents and supports lossless incremental updates. The system has been deployed in LinkedIn’s Hiring Assistant and serves production hiring workflows.
    \item \textbf{Hierarchical, tree-based memory representation:} We propose a tree-structured memory topology with multi-view memory representation, enabling massively parallel indexing, privacy-preserving retrieval scoping, and low-latency serving.
    \item \textbf{Adaptive ingestion via workload feedback:} We design a lightweight adaptation loop that mines historical query patterns to continuously refine memory content and improve retrieval quality and efficiency.
\end{itemize}

The remainder of the paper is organized as follows. Section \ref{ref_section_related_work} reviews related research and industrial practices in agent memory. Section \ref{ref_section_method} presents the design and methodology of \hltm. Section \ref{ref_section_experiments} provides a comparative evaluation against existing memory techniques. Section \ref{ref_production_use_case} describes \hltm's production use case in LinkedIn's Hiring Assistant. Finally, Section \ref{ref_section_conclusion} concludes the paper and outlines directions for future work.

\section{Related Work}
\label{ref_section_related_work}

Recent work has substantially improved agents' long-context processing by introducing explicit memory mechanisms. MemGPT~\cite{packer2023memgpt} and MemOS~\cite{li2025memos} adopt an OS-inspired design that swaps information between the limited context window and external memory. Mem0~\cite{chhikara2025mem0} stores discrete memory items and retrieves them primarily via semantic search. MemoryBank~\cite{zhong2024memorybank} maintains timestamped interaction memories with time-based decay to emphasize recency and salience. ReadAgent~\cite{lee2024human} compresses long documents into compact gist memories with on-demand lookup of supporting passages. SCM~\cite{wang2023scm} equips agents with a controller that adaptively decides when to store, update, or retrieve from a stream of past interactions.

Beyond unstructured memory stores, recent studies explore structured representations to improve organization and retrieval. GraphRAG~\cite{edge2024local} builds a knowledge graph and leverages community-based summarization to better capture entity-centric dependencies. HippoRAG~\cite{jimenez2024hipporag} draws inspiration from hippocampal indexing, integrating graph-based retrieval with personalized PageRank for multi-hop reasoning and continual knowledge integration. A-Mem~\cite{xu2025mem} uses Zettelkasten-style atomic notes with dynamic linking to support autonomous memory evolution, while G-Memory~\cite{zhang2025g} generalizes this direction to multi-agent settings with hierarchical graph structures that model both reusable insights and collaboration trajectories. In parallel, RAPTOR~\cite{sarthi2024raptor} and TreeRAG~\cite{tao2025treerag} leverage hierarchical structures for long-context retrieval, combining multi-level organization with structure-aware access to improve downstream QA. MemTree~\cite{rezazadeh2024isolated} complements this line with an incrementally updated memory tree that maintains node-level summaries and embeddings as agent knowledge evolves.

Despite these advances, enterprise deployment remains challenging. Many methods lack principled memory partitioning aligned with access boundaries, raising privacy and scope-isolation risks. Graph-based approaches often require extensive LLM usage during indexing and incur online latency, limiting scalability~\cite{zhao20252graphrag}. Tree-style methods frequently depend on the LLM's world knowledge for aggregation, which can weaken domain-specific controllability~\cite{maynez2020faithfulness}. Finally, most systems lack robust citation practices and provenance tracking, reducing transparency and debuggability in production~\cite{cheng2025ragtrace}.

\section{Method}
\label{ref_section_method}

\begin{figure*}[!t]
  \centering
  \includegraphics[width=\textwidth]{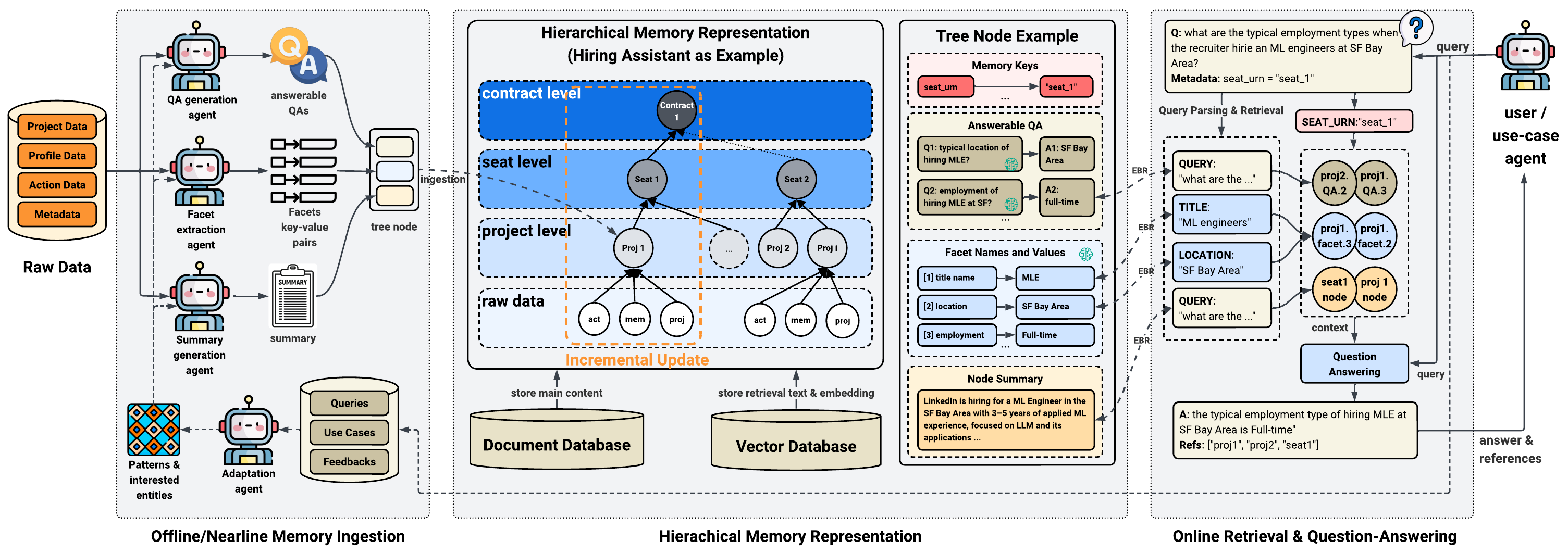}
  \caption{Overview of \hltm. Ingestion agents convert raw data into multi-view representations (left) and organize them in a schema-aligned memory tree (mid). At inference, relevant nodes are retrieved to generate answers (right).}
  \Description{Architecture diagram of the \hltm framework. Raw documents at the bottom feed into leaf memory nodes that are organized into a schema-aligned tree following business ownership (projects, recruiter seats, accounts). Each node stores three memory views: facet key--value pairs, answerable QA pairs, and summaries. Memory representations are aggregated bottom-up from children to parents. At query time, an identity-scoped subtree is selected and multi-signal retrieval over the three views supplies context to the LLM for answer generation.}
  \label{fig:overview_of_hltm}
\end{figure*}

\subsection{Overview}
We propose \hltm, a hierarchical long-term semantic memory that organizes information in a tree: higher-level nodes capture broad patterns, while lower-level nodes store fine-grained, entity-specific information. Together, these levels support both generic and highly targeted queries within a unified structure (Figure~\ref{fig:overview_of_hltm}). In production, the tree topology aligns with business scope and access-control boundaries, enabling privacy enforcement by restricting retrieval to appropriate subtrees (Section~\ref{ref_subsection_tree_topology_construction}). Node contents are built offline via data-driven preference extraction and hierarchical aggregation (Sections~\ref{ref_subsection_nodewise_knowledge_representation}--\ref{ref_subsection_hierarchical_knowledge_aggregation}), and are queried at serving time using hybrid embedding-based retrieval and answered via in-context learning (Section~\ref{ref_subsubsection_context_retrieval_via_hybrid_node_retrieval}). \hltm also supports adaptation to evolving query patterns and \emph{lossless} incremental updates, enabling nearline freshness at low indexing cost (Sections~\ref{ref_subsection_adaption_via_query_pattern_analysis}--\ref{subsection_incremental_update}).

\subsection{Schema-aligned Hierarchical Topology}
\label{ref_subsection_tree_topology_construction}

Most hierarchical memory systems adopt \emph{semantic hierarchies}, clustering embeddings of text segments into progressively coarser nodes~\cite{sarthi2024raptor,tao2025treerag}. While effective for organizing free-form text, this approach is ill-suited to multi-tenant enterprise privacy requirements: clustering can mix content across business scopes, creating privacy and compliance risks under strict isolation and access-control policies~\cite{zeng2024good,jiang2024rag}. Semantic hierarchies are also brittle under incremental updates: new data often triggers re-clustering and topology changes, while keeping the topology fixed can lead to quality drift and periodic full re-indexing.

\emph{Hierarchical schemas} are ubiquitous in enterprise data (ownership chains), documents (structural nesting), and conversations (sessions and messages). \hltm exploits this by constructing a \emph{schema-aligned hierarchy} whose topology is dictated by the enterprise data model and ownership boundaries. In the hiring domain, for example, leaf nodes represent fine-grained work units (e.g., hiring projects), intermediate nodes represent recruiter seats, and root nodes represent organizational accounts. We formalize this as a tree index $\mathcal{T} = (\mathcal{V}, \mathcal{E})$, where $\mathcal{V}$ contains business entities and edges $\mathcal{E}$ follow schema-defined parent--child relations. Each node $v \in \mathcal{V}$ owns the memory for its entity, and retrieval is restricted to subtrees consistent with the query scope and access-control policy.

This schema-aligned design provides three deployment-critical benefits:
(i) \textbf{privacy-aware scoping} via subtree-restricted retrieval, enabling per-entity isolation and controlled sharing;
(ii) \textbf{localized incremental updates} that touch only affected leaves and their ancestor nodes, reducing update cost from $O(|\mathcal{V}|)$ to the tree height; and
(iii) \textbf{topology stability} over time, avoiding drift and periodic full re-indexing commonly seen in clustering-based hierarchies.

\subsection{Memory Representation}
\label{ref_subsection_nodewise_knowledge_representation}
Rather than storing raw data at each node, we transform node content $D_v$ for node $v \in \mathcal{V}$ into a compact, multi-view memory representation $M_v$:
\[M_v = \big(F_v, Q_v, S_v\big)\]
where $F_v$, $Q_v$, and $S_v$ denote the facet, QA, and summary views at node $v \in \mathcal{V}$, respectively. The full prompt templates for the three agents are provided in Appendix~\ref{appendix:treemem_indexing_prompts}.
\begin{itemize}
    \item \textbf{Facet (Key--Value) Representation.}
    Key--value pairs provide a structured and interpretable view of salient facts, making them easy to index, filter, and retrieve. We prompt an LLM to extract a set of facet pairs $F_v = \{(\mathrm{key}, \mathrm{value})\}$ from $D_v$ and store them in a document database. To support semantic retrieval, we linearize each pair (e.g., \texttt{location: SF Bay Area}) and compute an embedding for indexing.

    \item \textbf{Answerable-QA Representation.}
    Question--answer pairs align with how users naturally phrase queries, so embedded questions often lie closer to user queries in embedding space, improving retrieval accuracy. They also enable a ``think-fast'' serving mode by precomputing answerable questions offline, minimizing online computation. Concretely, we prompt an LLM to generate answerable QA pairs $Q_v = \{(q,a)\}$ from the raw data $D_v$ and store them in a document database. To support semantic retrieval, we embed and index each question in a vector store.

    \item \textbf{Summary Representation.}
    Some knowledge is inherently unstructured (e.g., nuanced narratives or multi-step logic) and is not naturally captured by key--value or QA formats; therefore, we use natural-language summarization as the default representation. Specifically, we prompt an LLM to first compress $D_v$ into a detailed paragraph $S_v$ that preserves key information, and then distill it into a single-sentence summary to compute embeddings for vector-store indexing.
\end{itemize}

\subsection{Hierarchical Memory Aggregation}
\label{ref_subsection_hierarchical_knowledge_aggregation}
Summarization-style queries that ask for a summary or synthesis over a broad scope typically require aggregating evidence from multiple leaf nodes~\cite{edge2024local}. A straightforward strategy is to retrieve a large set of relevant leaf-level documents at query time and ask the LLM to summarize them on the fly; however, this approach is often impractical for three reasons: (i) \textbf{high online cost}, because summarizing detail-heavy evidence is both token-intensive and latency-critical; (ii) the \textbf{needle-in-a-haystack effect}, where relevant signals are drowned out by irrelevant details, increasing the likelihood of omissions and factual drift; and (iii) \textbf{embedding-space misalignment}, where broad summarization queries may not align well with leaf-level embeddings, leading to poor recall and unstable retrieval quality.

To address these challenges, \hltm performs \emph{offline hierarchical aggregation}, composing child memories into broader-scope parent representations in a bottom-up manner. Aggregation follows the parent--child relations defined by the tree topology (Section~\ref{ref_subsection_tree_topology_construction}) and is applied recursively up to the root. For each parent node $p \in \mathcal{V}$, we collect its children's memory representations $\{M_c\}$ as input and prompt an LLM agent to generate parent-level representations in the same multi-view form. Specifically, we guide the LLM to (i) group semantically similar evidence across children, (ii) merge redundant content and reconcile conflicts, and (iii) optionally prune low-salience details that rarely recur across children.

\subsection{Memory Retrieval and Answer Generation}
\label{ref_subsubsection_context_retrieval_via_hybrid_node_retrieval}
\hltm performs \emph{identity-scoped multi-signal retrieval} to efficiently retrieve privacy-compliant memory content for a given user query. Building on RAPTOR’s collapsed-tree paradigm~\cite{sarthi2024raptor}, which collapses nodes across levels into a single candidate pool for retrieval, we introduce two production-critical extensions: (i) \textbf{identity-scoped subtree filtering} to enforce enterprise access boundaries, and (ii) \textbf{multi-signal retrieval} to robustly surface relevant memories under diverse query formulations.

\subsubsection{Identity-scoped Subtree Filtering}
Each production query is issued under an explicit identity scope that maps to a node $v$ in our schema-aligned hierarchy (Section~\ref{ref_subsection_tree_topology_construction}). We enforce this scope via \emph{hard filtering}: retrieval candidates are restricted to the subtree $\mathcal{T}_{v}$ rooted at node $v$, guaranteeing strict tenant isolation. For example, a recruiter-scoped query may access only the recruiter node and its descendants; nodes owned by other recruiters are excluded. Formally,
\[
\mathcal{T}_v = \mathcal{T}\big[\{v\} \cup \mathrm{Desc}(v)\big]
\]

\subsubsection{Multi-signal Retrieval}
\label{ref_subsubsection_multi_dimensional_retrieval}
Within the privacy-scoped subtree $\mathcal{T}_{v}$, we build on RAPTOR’s collapsed-tree retrieval method~\cite{sarthi2024raptor} and extend it with multi-retriever scoring. Rather than relying on a single similarity signal, we rank candidates using three complementary retrievers tied to the node representations in Section~\ref{ref_subsection_nodewise_knowledge_representation}: the facet-based retriever capturing structured constraints, the answerable-QA retriever capturing query-like intents, and the summary-based retriever capturing free-form semantics. Each retriever produces a relevance score, and we return the top-$k$ memories per view as retrieval context.

\begin{itemize}

\item \textbf{Facet-based Retrieval:}
We enable fine-grained, schema-aware matching by parsing the query $q$ into a set of facet key--value pairs $F_q$. Each pair is then linearized into a micro-query (``key: $\mathrm{val}$'') and compared against the offline-constructed facet strings in $F_v$ for each node $v$ in embedding space. For example, the query \texttt{``typical workplace for hiring software engineers in the San Francisco Bay Area?''} is decomposed into two micro-queries: \texttt{``title: software engineer''} and \texttt{``location: San Francisco Bay Area''}. We aggregate similarities into a query--node score $s_{\text{facet}}(q,v)$ by averaging over the top-$k$ most similar node facets per micro-query, preventing irrelevant facet pairs from diluting the score, and return the facet sets $F_k(q)$ from the top-$k_{\text{facet}}$ nodes.

\hfill\begin{minipage}{\linewidth}
\[s_{\text{facet}}(q,v)
= \frac{1}{k\,|F_q|} \sum_{f \in F_q}
\sum_{f' \in \mathrm{TopK}(f,\, F_v)}
\cos\,\bigl(\mathrm{emb}(f), \mathrm{emb}(f')\bigr)\]
\end{minipage}

\item \textbf{Answerable-QA Retrieval:}
We match the query $q$ against a node’s offline-generated answerable question set $Q_v$. Specifically, we embed $q$ and each $q' \in Q_v$, compute cosine similarities, and score node $v$ by the best-matching question $s_{\text{QA}}(q,v)$. We then return the top-$k_{\text{QA}}$ answerable question--answer pairs $Q_k(q)$.

\hfill\begin{minipage}{\linewidth}
\[s_{\text{QA}}(q,v) = \max_{q' \in Q_v} \cos\,\bigl(\mathrm{emb}(q),\,\mathrm{emb}(q')\bigr)\]
\end{minipage}

\item \textbf{Summary-based Retrieval:}
We embed the query $q$ and each node’s offline-generated summary view $S_v$, rank nodes by cosine similarity, and return the top-$k_{\text{summary}}$ summaries $S_k(q)$.

\hfill\begin{minipage}{\linewidth}
\[
s_{\text{summary}}(q,v)
= \cos\,\bigl(\mathrm{emb}(q),\,\mathrm{emb}(S_v)\bigr)
\]
\end{minipage}

\end{itemize}

\subsubsection{Answer Generation with Retrieved Context}
\label{ref_question_answering_with_in_context_learning}
After retrieval, we apply a standard in-context learning prompt that conditions the LLM on the top-ranked memory contents as context. In addition to the final answer, the system also returns \emph{references}, namely the IDs of the retrieved memory tree nodes used as grounding evidence (Figure~\ref{fig:LiHA_UI}), enabling provenance tracking and downstream inspection. The prompt template is provided in Appendix~\ref{appendix_prompt_question_answering}.

\[
\mathrm{answer}, \mathrm{references} = \mathrm{LLM}\bigl(q,\, F_{k}(q)\cup Q_{k}(q)\cup S_{k}(q) \bigr)
\]

\subsection{Adaptation via Query Pattern Analysis}
\label{ref_subsection_adaption_via_query_pattern_analysis}
To remain adaptive under continuously shifting production query patterns and an evolving document corpus, we introduce a periodic adaptation step that distills historical query workloads into signals for memory construction, rather than relying solely on an LLM’s world knowledge to determine what is important to memorize. Over a sliding window of past queries, the module extracts (i) frequently recurring query patterns and (ii) salient facet names. The query patterns act as priors for the Answerable-QA Generation Agent, guiding it to synthesize answerable questions from node raw data $D_v$, while facet names provide optional hints for the Facet Extraction Agent and Summary Agents to emphasize coverage of high-impact attributes during memory construction and tree aggregation. To reduce overfitting, we apply minimum-support thresholds (i.e., only patterns/facets supported by multiple queries are retained); for sensitive production settings, we optionally require human review before deploying updates to the indexing pipeline.

\subsection{Lossless Incremental Nearline Indexing}
\label{subsection_incremental_update}
\begin{figure}
    \centering
    \includegraphics[width=1\linewidth]{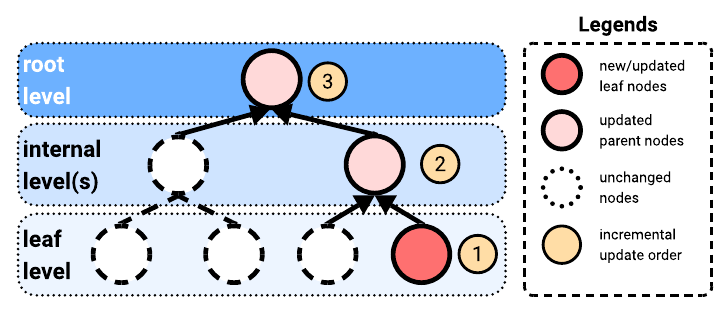}
    \caption{Lossless incremental nearline indexing in \hltm: updated leaves are re-indexed and their aggregates propagated along ancestor paths, while unaffected subtrees remain unchanged.}
    \Description{A diagram of lossless incremental nearline indexing in \hltm: updated leaf nodes are re-indexed, and updates are propagated along ancestor paths, while unaffected subtrees remain unchanged.}
    \label{fig:incremental_update}
\end{figure}

Data freshness is critical for user-facing agents; for example, recruiter memories in LinkedIn’s Hiring Assistant should reflect new interactions within minutes. Full indexing that re-runs all LLM-based summarization and embedding steps is too expensive; incremental updates such as GraphRAG~\cite{edge2024local} are typically \emph{lossy} and still require periodic full re-indexing.

\hltm natively supports lossless incremental nearline indexing, keeping large-scale memories fresh without full re-indexing. As shown in Figure~\ref{fig:incremental_update}, we maintain a set $V^{\star}$ of leaf nodes newly created or modified since the last cycle; for each $v \in V^{\star}$, we recompute its memory representation $M_v$ as discussed in Section~\ref{ref_subsection_nodewise_knowledge_representation}, and then update only its ancestor path $\{u \in V \mid v \preceq u\}$ up to the root. Because each parent is always recomputed from its latest children, the resulting tree is identical to a full rebuild over the latest data snapshot. This enables nearline update windows of a few minutes that are substantially cheaper than full re-indexing while preserving the semantics of a fresh global rebuild. We evaluated the gain from incremental indexing in production in Appendix~\ref{appendix:incremental_indexing}.

\section{Experiments}
\label{ref_section_experiments}

\subsection{Dataset}
\paragraph{LinkedIn’s Hiring Assistant Dataset} We constructed a human-labeled benchmark dataset derived from LinkedIn’s Hiring Assistant usage logs. The dataset spans diverse hiring domains, with gold reference answers curated by domain experts. Each data point contains (i) \textbf{raw hiring documents}, including a recruiter’s historical hiring projects, within-project interactions on candidates, and anonymized candidate and project data; (ii) \textbf{memory queries} issued by the Hiring Assistant, spanning both \emph{summary-style} queries that aggregate signals across projects to infer preferences (e.g., typical titles, locations, and qualifications) and \emph{retrieval-style} queries that request targeted projects or evidence snippets for downstream reasoning; and (iii) \textbf{gold answers}, which are personalized to the underlying hiring documents. To improve label reliability, each answer is independently annotated by at least three annotators, and disagreements are resolved via majority voting. 
\paragraph{LongMemEval-s Dataset} To assess \hltm's performance beyond the hiring domain, we additionally evaluate on LongMemEval-s~\cite{wu2024longmemeval}, a publicly available benchmark designed to stress-test long-term memory in conversational AI systems.

\begin{table}[H]
  \centering
  \caption{Dataset summary statistics.}
  \label{tab:dataset_stats}
  \small
  \setlength{\tabcolsep}{6pt}
  \renewcommand{\arraystretch}{1.05}
  \begin{tabular}{@{}l c c@{}}
    \toprule
    \textbf{Dataset Metrics} & \textbf{LinkedIn Dataset} & \textbf{LongMemEval-s} \\
    \midrule
    \# Queries                     & 1,341  & 500    \\
    \quad -- Retrieval queries     & 473    & ---    \\
    \quad -- Summarization queries & 868    & ---    \\
    Avg.\ history length (tokens)   & $\approx$40K & $\approx$103K \\
    Total history length (tokens)                   & $\approx$53M & $\approx$51M  \\
    \bottomrule
  \end{tabular}
\end{table}

\subsection{Evaluation Metrics}

\subsubsection{Performance Metrics}
\label{ref_subsubsection_performance_metrics}
\paragraph{Lexical Similarity Metrics.}
Prior work in conversational AI and agent memory evaluation~\cite{chhikara2025mem0, xu2025mem, liu2026simplemem} commonly reports lexical overlap metrics. We report token-level F1 to balance spurious content and missing coverage, and BLEU-1 to provide a complementary, precision-oriented lexical similarity signal.

\paragraph{Semantic Correctness Metrics.}
To evaluate answer correctness beyond surface-form overlap, we additionally employ an LLM-as-a-judge score. The judge is given the query, the system-generated answer, and the gold reference answer, and scores factual consistency with the reference by checking key constraints such as numbers and named entities. The judge is instructed to rely only on the gold reference and outputs a scalar score in $[0,1]$ along with a brief rationale. To validate reliability, we conducted a human alignment study to ensure the LLM judge agrees with expert judgment. Specifically, we collected approximately 200 query--answer triples independently annotated by 2--4 domain experts; inter-annotator agreement (Quadratic Weighted Cohen's $\kappa$) exceeded 0.8, and the LLM judge benchmarked against majority-vote consensus labels achieved the same threshold, confirming strong alignment with human judgment.

\subsubsection{Deployment Metrics}
\label{ref_subsubsection_deployment_metrics}
\paragraph{Latency.}
We measure both \emph{offline indexing latency} and \emph{online query latency}. Offline indexing latency indicates whether a method can refresh and maintain memory at production scale (e.g., millions of documents ingested daily for LinkedIn’s Hiring Assistant). Online query latency directly affects end-user experience in interactive, chat-based workflows. When a method supports parallelism, we run indexing and retrieval with a fixed concurrency setting to reflect realistic deployment conditions.

\paragraph{LLM Invocation Counts and Token Consumption.}
We track both LLM call counts and token usage, since together they largely determine serving cost. Our accounting covers all LLM interactions during the query phase, including query rewriting, planning, retrieval-time processing, and final answer generation. We compute token usage with a consistent tokenizer (\texttt{o200k\_base}).

\subsection{Baselines}
To comprehensively evaluate \hltm, we compare against both basic baselines and advanced RAG/memory methods. We hold the backbone LLM and embedding model fixed across methods, and evaluate all systems under the same serving setup and context-window constraints.

\subsubsection{Basic Baselines}
We include three simple yet strong baselines, following the evaluation practice of Mem0~\cite{chhikara2025mem0}, to isolate the value of an external memory module. (i) \textbf{Conventional retrieval-augmented generation (RAG)}~\cite{lewis2020retrieval} chunks documents into fixed-length segments, embeds each chunk, and retrieves the top chunks by semantic similarity as context for question answering. (ii) \textbf{Full-context prompting} feeds the full available history (or the maximal portion that fits within the LLM context window) directly into the LLM, providing an architecture-free baseline. (iii) \textbf{Schema filter} parses each document into a structured table at ingestion time and applies query parsing at retrieval time to filter matching documents, which are then passed to the LLM for answer synthesis. This baseline directly exploits domain schema knowledge, isolating the contribution of \hltm's hierarchical aggregation and multi-view memory.

\subsubsection{Advanced Baselines}
We further compare against representative advanced RAG and memory systems that introduce explicit structure beyond flat retrieval.
(i) \textbf{Graph-based methods} represent memory as nodes and edges, enabling entity-centric organization and multi-hop retrieval; we include GraphRAG~\cite{edge2024local}, HippoRAG~\cite{jimenez2024hipporag}, A-Mem~\cite{xu2025mem}, and SimpleMem~\cite{liu2026simplemem}. (ii) \textbf{Tree-based methods} build hierarchical indices via recursive clustering and summarization to balance coverage and retrieval efficiency; we evaluate RAPTOR~\cite{sarthi2024raptor}, and leave TreeRAG~\cite{tao2025treerag} and MemTree~\cite{rezazadeh2024isolated} for future comparison once runnable implementations are publicly available. (iii) \textbf{Other common baselines} include ReadAgent~\cite{lee2024human} and Mem0~\cite{chhikara2025mem0}, which serve as widely used reference points for long-context retrieval and memory-augmented generation.

\subsection{Implementation Details}
\hltm was originally deployed in LinkedIn’s production environment with PySpark-based offline ingestion and in-house document/vector stores. For this study, we re-implemented \hltm as a standalone Python library for benchmarking. All methods run on Python~3.12 and access LLMs via the Azure OpenAI Service API. We use \texttt{text-embedding-3-large} for embeddings, \texttt{GPT-4o mini} as the primary LLM, and four widely-used open-source language models ranging from 27B to 106B parameters. We report \texttt{GPT-4o mini} results in the main paper and provide open-source models' results in Appendix~\ref{appendix:detailed_experiment_results}. Baselines use default open-source settings unless noted, and all results are averaged over three runs.

\subsection{Experiment Result Analysis}
\label{ref_subsection_experiment_result_analysis}
\begin{table*}[t!]
\centering
\small
\setlength{\tabcolsep}{5pt}
\caption{LinkedIn dataset answer performance; Values are mean $\pm$ SEM; $\uparrow$ indicates higher is better.}
\label{tab:main_results_two_query_types}
\begin{tabular}{l c c c | c c c}
\toprule
& \multicolumn{3}{c|}{\textbf{Summary-style Queries}}
& \multicolumn{3}{c}{\textbf{Retrieval-style Queries}} \\
\cmidrule(lr){2-4} \cmidrule(lr){5-7}
\textbf{Methods}
& \textbf{Token-F1 $\uparrow$}
& \textbf{BLEU-1 $\uparrow$}
& \textbf{Correctness $\uparrow$}
& \textbf{Precision $\uparrow$}
& \textbf{Recall $\uparrow$}
& \textbf{F1 $\uparrow$}\\
\midrule
\textsc{Full-context}
& $0.478 \pm 0.007$ & $0.348 \pm 0.007$ & $0.791 \pm 0.009$
& $0.465 \pm 0.017$ & $0.849 \pm 0.015$ & $0.544 \pm 0.015$ \\
\textsc{RAG}
& $0.534 \pm 0.008$ & $0.409 \pm 0.008$ & $0.770 \pm 0.011$
& $0.557 \pm 0.017$ & $0.762 \pm 0.018$ & $0.604 \pm 0.016$ \\
\textsc{Schema Filter}
& $0.352 \pm 0.011$ & $0.269 \pm 0.010$ & $0.513 \pm 0.016$
& $0.541 \pm 0.022$ & $0.575 \pm 0.022$ & $0.546 \pm 0.021$ \\
\midrule
\textsc{A-Mem}
& $0.523 \pm 0.007$ & $0.391 \pm 0.007$ & $0.779 \pm 0.010$
& $0.498 \pm 0.017$ & $0.719 \pm 0.019$ & $0.549 \pm 0.017$ \\
\textsc{HippoRAG}
& $0.453 \pm 0.005$ & $0.329 \pm 0.005$ & $0.833 \pm 0.010$
& $0.515 \pm 0.018$ & $0.718 \pm 0.019$ & $0.560 \pm 0.017$ \\
\textsc{RAPTOR}
& $0.539 \pm 0.008$ & $0.423 \pm 0.008$ & $0.715 \pm 0.012$
& $0.332 \pm 0.018$ & $0.425 \pm 0.021$ & $0.348 \pm 0.018$ \\
\textsc{GraphRAG}
& $0.114 \pm 0.003$ & $0.074 \pm 0.002$ & $0.381 \pm 0.013$
& $0.082 \pm 0.011$ & $0.112 \pm 0.014$ & $0.084 \pm 0.011$ \\
\textsc{SimpleMem}
& $0.150 \pm 0.006$ & $0.032 \pm 0.003$ & $0.518 \pm 0.015$
& $0.445 \pm 0.019$ & $0.570 \pm 0.022$ & $0.470 \pm 0.020$ \\
\textsc{Mem0}
& $0.470 \pm 0.008$ & $0.372 \pm 0.008$ & $0.521 \pm 0.013$
& $0.376 \pm 0.020$ & $0.412 \pm 0.022$ & $0.369 \pm 0.020$ \\
\textsc{ReadAgent}
& $0.473 \pm 0.007$ & $0.347 \pm 0.008$ & $0.796 \pm 0.009$
& $0.542 \pm 0.016$ & $0.861 \pm 0.014$ & $0.617 \pm 0.015$ \\
\specialrule{\lightrulewidth}{0pt}{0pt}
\rowcolor{gray!15}\hltm (ours)
& $\mathbf{0.724 \pm 0.008}$
& $\mathbf{0.588 \pm 0.008}$
& $\mathbf{0.892 \pm 0.007}$
& $\mathbf{0.761 \pm 0.016}$
& $\mathbf{0.874 \pm 0.014}$
& $\mathbf{0.782 \pm 0.014}$ \\
\specialrule{\heavyrulewidth}{0pt}{0pt}
\end{tabular}
\end{table*}

\begin{table*}[t!]
\centering
\small
\setlength{\tabcolsep}{4pt}
\caption{LongMemEval-s answer accuracy, overall and per query type; Values are mean $\pm$ SEM.}
\label{tab:longmemeval_main_results_two_query_types}
\begin{tabular}{lccccccc}
\toprule
  &
  & \multicolumn{6}{c}{\textbf{By Question Type}} \\
\cmidrule(lr){3-8}
\textbf{Methods} & \textbf{Overall}
  & \textbf{know-update} & \textbf{multi-sess} & \textbf{single-sess-asst} & \textbf{single-sess-pref} & \textbf{single-sess-user} & \textbf{temp-reason} \\
\midrule
\textsc{Full-context} & $0.494 \pm 0.022$ & $0.679 \pm 0.053$ & $0.406 \pm 0.043$ & $0.607 \pm 0.065$ & $0.067 \pm 0.046$ & $0.814 \pm 0.046$ & $0.353 \pm 0.041$ \\
\textsc{RAG} & $0.435 \pm 0.017$ & $0.415 \pm 0.034$ & $0.207 \pm 0.024$ & $0.923 \pm 0.022$ & $0.528 \pm 0.069$ & $0.783 \pm 0.031$ & $0.264 \pm 0.027$ \\
\cmidrule(lr){1-8}
\textsc{A-Mem} & $0.716 \pm 0.020$ & $0.846 \pm 0.040$ & $0.556 \pm 0.043$ & $0.946 \pm 0.030$ & $0.567 \pm 0.090$ & $0.886 \pm 0.038$ & $0.647 \pm 0.041$ \\
\textsc{HippoRAG} & $0.578 \pm 0.022$ & $0.667 \pm 0.053$ & $0.534 \pm 0.043$ & $0.946 \pm 0.030$ & $0.100 \pm 0.054$ & $0.829 \pm 0.045$ & $0.391 \pm 0.042$ \\
\textsc{RAPTOR} & $0.530 \pm 0.022$ & $0.577 \pm 0.056$ & $0.376 \pm 0.042$ & $0.821 \pm 0.051$ & $0.433 \pm 0.091$ & $0.871 \pm 0.040$ & $0.376 \pm 0.042$ \\
\textsc{Mem0} & $0.462 \pm 0.023$ & $0.564 \pm 0.057$ & $0.444 \pm 0.043$ & $0.071 \pm 0.034$ & $0.400 \pm 0.089$ & $0.871 \pm 0.040$ & $0.383 \pm 0.042$ \\
\textsc{ReadAgent} & $0.480 \pm 0.023$ & $0.538 \pm 0.057$ & $0.316 \pm 0.040$ & $0.821 \pm 0.051$ & $0.467 \pm 0.091$ & $0.771 \pm 0.050$ & $0.316 \pm 0.040$ \\
\cmidrule(lr){1-8}
\cellcolor{gray!15}\hltm (ours) & \cellcolor{gray!15}$\mathbf{0.778} \pm \mathbf{0.019}$ & \cellcolor{gray!15}$\mathbf{0.872} \pm \mathbf{0.037}$ & \cellcolor{gray!15}$\mathbf{0.632} \pm \mathbf{0.042}$ & \cellcolor{gray!15}$\mathbf{0.964} \pm \mathbf{0.025}$ & \cellcolor{gray!15}$\mathbf{0.667} \pm \mathbf{0.086}$ & \cellcolor{gray!15}$\mathbf{0.986} \pm \mathbf{0.014}$ & \cellcolor{gray!15}$\mathbf{0.707} \pm \mathbf{0.039}$ \\
\bottomrule
\end{tabular}
\end{table*}

\subsubsection{Overview}
Figure~\ref{fig:performance-vs-latency} shows that \hltm achieves the best quality--latency trade-off on our benchmark, delivering the highest correctness (0.892) at low query latency ($\approx$3\,s). Methods with similar latency (e.g., RAG) are substantially less accurate, whereas higher-performing baselines such as HippoRAG~\cite{jimenez2024hipporag} achieve improved correctness only at a substantially higher latency ($\approx$7\,s).

\subsubsection{Quality Analysis.}
Table~\ref{tab:main_results_two_query_types} reports results on LinkedIn's Hiring Assistant dataset using \texttt{GPT-4o mini}. \hltm consistently outperforms all ten baselines across both query types\footnote{Disclaimer: Results may vary in production environments or with different datasets.}. For summary-style queries, \hltm improves semantic correctness by more than 5\% over the strongest baseline, HippoRAG (0.833). For retrieval-style queries, \hltm improves retrieval F1 by more than 10\% over the best baseline, ReadAgent (0.617), demonstrating a favorable precision--recall trade-off. Both gains are statistically significant under a Wilcoxon signed-rank test ($p < 10^{-7}$), with Cliff's $\delta = +0.058$ for summary-style queries against HippoRAG and $\delta = +0.328$ for retrieval-style queries against ReadAgent. Table~\ref{tab_appendix:performance_metrics} in Appendix~\ref{appendix:detailed_experiment_results} further shows that these improvements are consistent across four open-source language models ranging from 27B to 106B parameters. Table~\ref{tab:longmemeval_main_results_two_query_types} reports results on LongMemEval-s, which probes long-term memory across six question types spanning knowledge updates, multi-session reasoning, single-session retrieval (assistant, user-preference, and user-information views), and temporal reasoning. \hltm again outperforms all baselines, improving overall correctness by 6\% over the strongest baseline, A-Mem (0.716), and achieves the best score on every one of the six question types. One of the largest gains appears on multi-session reasoning (0.632 vs.\ A-Mem 0.556), which most directly stresses cross-session aggregation that \hltm's hierarchical aggregation is designed to handle.

\subsubsection{Latency and Cost Analysis.}
\begin{figure}[H]
    \centering
    \includegraphics[width=1\linewidth]{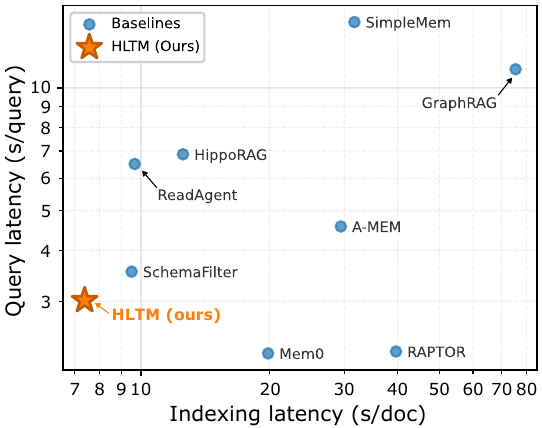}
    \caption{Query vs.\ indexing latency: \hltm advances the Pareto frontier.}
    \Description{Scatter plot of offline indexing latency versus online query latency across evaluated methods. \hltm lies on the Pareto frontier, achieving low query latency together with competitive indexing latency, dominating baselines that are slower on one or both axes.}
    \label{fig:query_indexing_latency_plot}
\end{figure}

Figure~\ref{fig:query_indexing_latency_plot} highlights the trade-off between offline indexing latency and online query latency. \hltm pushes the Pareto frontier by enabling parallel indexing across child nodes and reducing online latency through collapsed-tree retrieval. Token analysis further shows that \hltm reduces query-time token usage by at least 50\% relative to graph-based baselines, indicating more efficient context selection and compaction (Table~\ref{tab_appendix:deployment_metrics} in Appendix~\ref{appendix_subsection_deployment_metrics}).

\subsection{Ablation Study and Analysis}
We conduct an ablation study to quantify the contributions of tree aggregation, adaptation, and each memory representation. Table~\ref{tab:treemem_ablation_answer_quality} shows that disabling \textbf{tree aggregation} (i.e., restricting retrieval to leaf nodes only) causes a clear drop in summary-style correctness (0.892$\rightarrow$0.850; $\approx$4 points) and a large drop in retrieval-style F1 (0.782$\rightarrow$0.618; $\approx$16 points). Removing \textbf{adaptation} also degrades quality (correctness 0.892$\rightarrow$0.829; $\approx$6 points; retrieval F1 0.782$\rightarrow$0.593; $\approx$19 points), indicating that aligning aggregation with workload-specific query patterns remains important even when the core hierarchy is intact. Finally, single-representation variants suggest that \textbf{summary} memory contributes most to overall performance, while \textbf{facet} and \textbf{QA} memories provide complementary gains; combining all three yields the best correctness on summary-style queries and the best F1 on retrieval-style queries.

\begin{table*}[t!]
\centering
\small
\setlength{\tabcolsep}{5pt}
\caption{\hltm ablation results on LinkedIn dataset answer performance; Values are mean $\pm$ SEM; $\uparrow$ indicates higher is better.}
\label{tab:treemem_ablation_answer_quality}
\begin{tabular}{l c c c | c c c}
\toprule
& \multicolumn{3}{c|}{\textbf{Summary-style Queries}}
& \multicolumn{3}{c}{\textbf{Retrieval-style Queries}} \\
\cmidrule(lr){2-4} \cmidrule(lr){5-7}
\textbf{Methods}
& \textbf{Token-F1 $\uparrow$}
& \textbf{BLEU-1 $\uparrow$}
& \textbf{Correctness $\uparrow$}
& \textbf{Precision $\uparrow$}
& \textbf{Recall $\uparrow$}
& \textbf{F1 $\uparrow$}\\
\midrule
\hltm
& $\mathbf{0.724} \pm \mathbf{0.008}$
& $\mathbf{0.588} \pm \mathbf{0.008}$
& $\mathbf{0.892} \pm \mathbf{0.008}$
& $\mathbf{0.761} \pm \mathbf{0.016}$
& $0.874 \pm 0.014$
& $\mathbf{0.782} \pm \mathbf{0.014}$ \\
\midrule
\hltm (w/o aggregation) & $0.672 \pm 0.007$ & $0.525 \pm 0.007$ & $0.850 \pm 0.008$ & $0.568 \pm 0.018$ & $0.784 \pm 0.017$ & $0.618 \pm 0.016$ \\
\hltm (w/o adaptation)
& $0.689 \pm 0.008$
& $0.555 \pm 0.009$
& $0.829 \pm 0.010$
& $0.524 \pm 0.016$
& $0.811 \pm 0.016$
& $0.593 \pm 0.015$ \\
\midrule
\hltm (facet-only)
& $0.619 \pm 0.010$
& $0.502 \pm 0.009$
& $0.809 \pm 0.011$
& $0.552 \pm 0.018$
& $0.755 \pm 0.019$
& $0.599 \pm 0.017$ \\
\hltm (QA-only)
& $0.661 \pm 0.009$
& $0.547 \pm 0.010$
& $0.710 \pm 0.012$
& $0.642 \pm 0.019$
& $0.717 \pm 0.019$
& $0.646 \pm 0.018$ \\
\hltm (summary-only)
& $0.680 \pm 0.008$
& $0.537 \pm 0.008$
& $0.850 \pm 0.008$
& $0.684 \pm 0.017$
& $\mathbf{0.877} \pm \mathbf{0.014}$
& $0.730 \pm 0.015$ \\
\bottomrule
\end{tabular}
\end{table*}

\subsection{Privacy Discussion}
\label{ref_subsection_privacy}
Privacy is a first-class requirement for production memory systems. In \hltm, isolation is enforced structurally: each node is bound to an identity scope (e.g., recruiter, hiring project), and retrieval is restricted to the subtree rooted at that scope. Recruiter-scoped queries can access only that recruiter node and its owned projects; project-scoped queries are further constrained to a single project. This business-aligned hierarchy lets us \emph{index once} and \emph{query at multiple levels} without per-level re-indexing or global LLM routing, while preserving strict privacy isolation (Table \ref{tab:leakage_experiment} in Appendix \ref{section:privacy_isolation_experiments}).

\section{Production Use Case}
\label{ref_production_use_case}

\begin{figure}[tb!]
    \centering
    \includegraphics[width=0.99\linewidth]{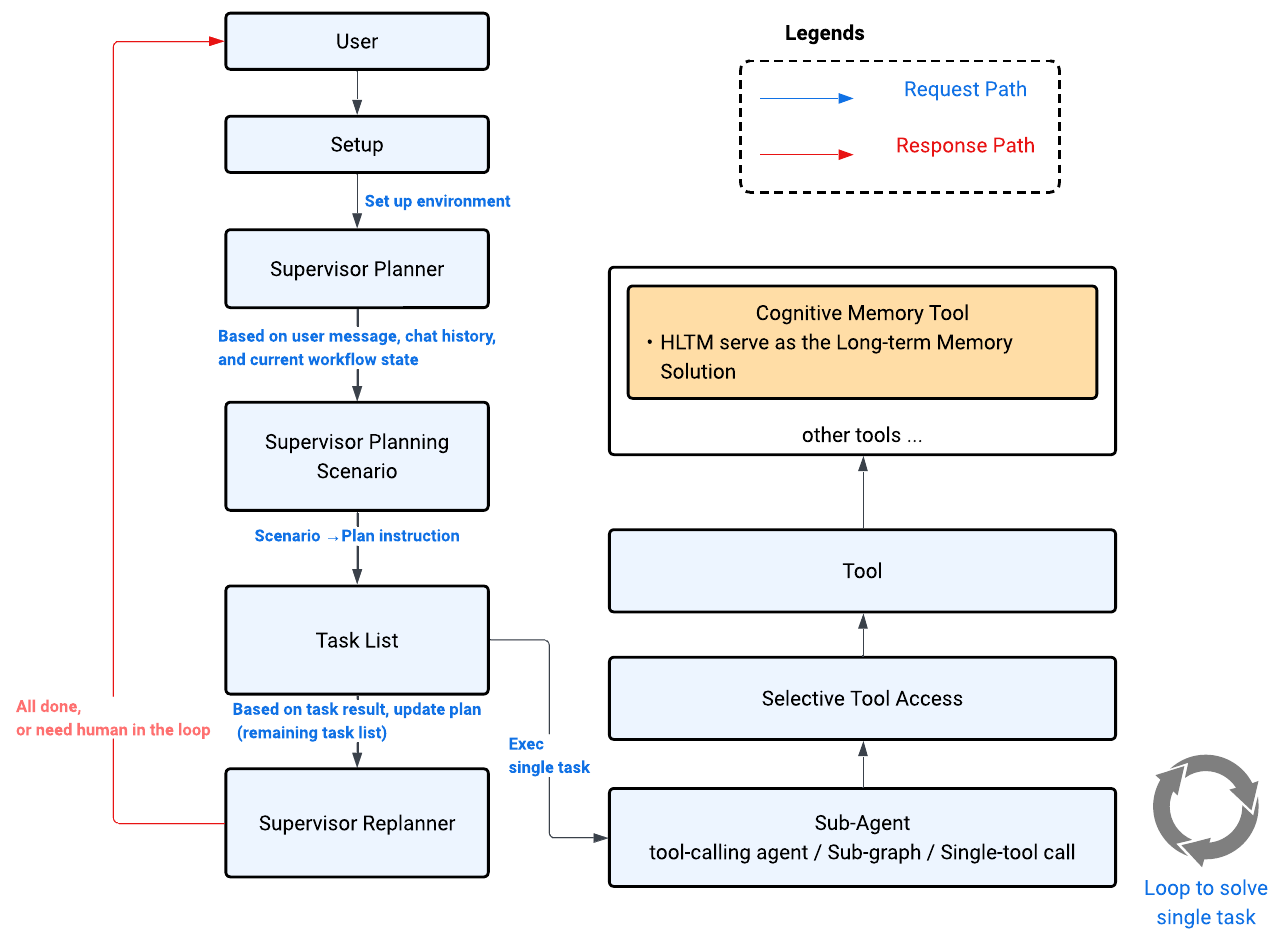}
    \caption{\hltm's Production Use Case in LinkedIn's Hiring Assistant.}
    \Description{System diagram of \hltm deployed inside LinkedIn's Hiring Assistant. A supervisor agent receives recruiter requests, plans actions, and dispatches to specialized sub-agents and tools, including the \hltm long-term memory service that returns recruiter-specific preference signals to personalize responses.}
    \label{fig:liha_diagram}
\end{figure}

LinkedIn's Hiring Assistant~\cite{gu2025hiringassistant} is an AI recruiting agent that
helps recruiters discover, evaluate, and engage candidates with greater speed and scale.
Architecturally, LinkedIn's Hiring Assistant is a plan-and-execute system centered on a supervisor agent that
interprets recruiter intent and orchestrates specialized sub-agents and tools
(Figure~\ref{fig:liha_diagram}).
\hltm has been fully ramped as the long-term semantic memory of LinkedIn's Hiring Assistant since early 2026 (Figure~\ref{fig:LiHA_UI}). Concretely, \hltm participates in the user-agent chat process by dynamically recalling each recruiter's long-term preferences from historical projects and
activities, enabling the assistant to personalize interactions without requiring recruiters to
re-specify requirements that can be inferred from past behavior.\footnote{Disclaimer: All memory data is scoped to the recruiter's environment and is not used for training content-generating AI models; customers retain full control over their stored memory with robust privacy and management options.}

We assess \hltm's online impact by analyzing user satisfaction across 3,000$+$ eligible production conversations from the initial rollout. The supervisor agent autonomously decides when to invoke the memory module based on conversational context, triggering it in over 40\% of sessions --- a strong signal that a large fraction of recruiter interactions benefit from long-term preference recall. Across 1,000$+$ recruiter seats, sessions in which \hltm was invoked show a \textbf{5--10 percentage-point reduction in negative feedback rate} during the hiring-calibration phase, statistically significant under a two-proportion $z$-test ($p < 0.01$). This improvement indicates that candidates and recommendations surfaced during calibration are of noticeably higher quality when the assistant leverages historical preference memory, demonstrating \hltm's tangible impact on real recruiter workflows at production scale.

\section{Conclusion}
\label{ref_section_conclusion}
We introduce \hltm, a hierarchical, schema-aligned long-term memory framework for industrial LLM agents, and demonstrate its end-to-end deployment in LinkedIn’s Hiring Assistant~\cite{gu2025hiringassistant}. \hltm organizes memories into a tree topology that mirrors real-world business scopes (Section~\ref{ref_subsection_tree_topology_construction}) and generates memory content via complementary multi-view representations with bottom-up hierarchical aggregation (Sections~\ref{ref_subsection_nodewise_knowledge_representation}–\ref{ref_subsection_hierarchical_knowledge_aggregation}). Experiments on production hiring data show that \hltm consistently outperforms strong RAG and memory baselines, with semantic correctness gains of more than 5\% on summary-style queries and F1 gains of more than 10\% on retrieval-style queries\footnote{Disclaimer: Results may vary in production environments or with different datasets.}, while maintaining favorable latency and token cost for real-time, chat-based workflows (Section~\ref{ref_subsection_experiment_result_analysis}). In production, \hltm supports large-scale indexing with freshness guarantees via lossless incremental ingestion (Section~\ref{subsection_incremental_update}) and provides privacy and observability through business-keyed, transparent nodes (Section~\ref{ref_subsection_privacy}), making it a practical foundation for industrial GenAI products. \hltm has been fully deployed in LinkedIn’s Hiring Assistant to support real recruiter hiring workflows (Section~\ref{ref_production_use_case}). Looking ahead, we plan to extend \hltm to richer cross-entity reasoning, multimodal signals, and additional agent use cases beyond hiring, and to explore tighter integration with agent planners and controllers for end-to-end adaptive systems.

\begin{acks}
This work was conducted by the Agent Platform team at LinkedIn. We thank our collaborators on the LinkedIn Talent Solutions team for building the Hiring Assistant product and for partnering with us to integrate \hltm as its long-term memory solution. We also thank our leadership team, especially Anuradha Krishnan, Deepak Agarwal, Balaji Krishnapuram, and Erran Berger for their guidance and support.
\end{acks}

\bibliographystyle{ACM-Reference-Format}
\bibliography{main}

@String{Computer = "{IEEE} Computer" }

@String{Springer = "Springer-Verlag" }

@article{packer2023memgpt,
  title={MemGPT: Towards LLMs as Operating Systems.},
  author={Packer, Charles and Fang, Vivian and Patil, Shishir\_G and Lin, Kevin and Wooders, Sarah and Gonzalez, Joseph\_E},
  year={2023},
  publisher={ArXiv}
}

@inproceedings{chhikara2025mem0,
  title={Mem0: Building production-ready ai agents with scalable long-term memory},
  author={Chhikara, Prateek and Khant, Dev and Aryan, Saket and Singh, Taranjeet and Yadav, Deshraj},
  booktitle={Proceedings of the 28th European Conference on Artificial Intelligence},
  pages={2993--3000},
  year={2025},
  doi={10.3233/FAIA251160}
}

@article{edge2024local,
  title={From local to global: A graph rag approach to query-focused summarization},
  author={Edge, Darren and Trinh, Ha and Cheng, Newman and Bradley, Joshua and Chao, Alex and Mody, Apurva and Truitt, Steven and Metropolitansky, Dasha and Ness, Robert Osazuwa and Larson, Jonathan},
  journal={arXiv preprint arXiv:2404.16130},
  year={2024}
}

@inproceedings{sarthi2024raptor,
  title={Raptor: Recursive abstractive processing for tree-organized retrieval},
  author={Sarthi, Parth and Abdullah, Salman and Tuli, Aditi and Khanna, Shubh and Goldie, Anna and Manning, Christopher D},
  booktitle={The Twelfth International Conference on Learning Representations},
  year={2024}
}

@article{rezazadeh2024isolated,
  title={From isolated conversations to hierarchical schemas: Dynamic tree memory representation for llms},
  author={Rezazadeh, Alireza and Li, Zichao and Wei, Wei and Bao, Yujia},
  journal={arXiv preprint arXiv:2410.14052},
  year={2024}
}

@article{jimenez2024hipporag,
  title={Hipporag: Neurobiologically inspired long-term memory for large language models},
  author={Jimenez Gutierrez, Bernal and Shu, Yiheng and Gu, Yu and Yasunaga, Michihiro and Su, Yu},
  journal={Advances in Neural Information Processing Systems},
  volume={37},
  pages={59532--59569},
  year={2024}
}

@inproceedings{wang2023scm,
  title={SCM: Enhancing large language model with self-controlled memory framework},
  author={Wang, Bing and Liang, Xinnian and Yang, Jian and Huang, Hui and Wu, Shuangzhi and Wu, Peihao and Lu, Lu and Ma, Zejun and Li, Zhoujun},
  booktitle={Proceedings of the 30th International Conference on Database Systems for Advanced Applications},
  year={2025}
}

@inproceedings{zhong2024memorybank,
  title={Memorybank: Enhancing large language models with long-term memory},
  author={Zhong, Wanjun and Guo, Lianghong and Gao, Qiqi and Ye, He and Wang, Yanlin},
  booktitle={Proceedings of the AAAI Conference on Artificial Intelligence},
  volume={38},
  number={17},
  pages={19724--19731},
  year={2024}
}

@inproceedings{lee2024human,
  title={A human-inspired reading agent with gist memory of very long contexts},
  author={Lee, Kuang-Huei and Chen, Xinyun and Furuta, Hiroki and Canny, John and Fischer, Ian},
  booktitle={Proceedings of the 41st International Conference on Machine Learning},
  series={Proceedings of Machine Learning Research},
  volume={235},
  pages={26396--26415},
  year={2024}
}

@inproceedings{xu2025mem,
  title={A-Mem: Agentic memory for LLM agents},
  author={Xu, Wujiang and Liang, Zujie and Mei, Kai and Gao, Hang and Tan, Juntao and Zhang, Yongfeng},
  booktitle={Advances in Neural Information Processing Systems},
  year={2025}
}

@article{zhang2025g,
  title={G-Memory: Tracing Hierarchical Memory for Multi-Agent Systems},
  author={Zhang, Guibin and Fu, Muxin and Wan, Guancheng and Yu, Miao and Wang, Kun and Yan, Shuicheng},
  journal={arXiv preprint arXiv:2506.07398},
  year={2025}
}

@inproceedings{tao2025treerag,
  title={Treerag: Unleashing the power of hierarchical storage for enhanced knowledge retrieval in long documents},
  author={Tao, Wenyu and Xing, Xiaofen and Chen, Yirong and Huang, Linyi and Xu, Xiangmin},
  booktitle={Findings of the Association for Computational Linguistics: ACL 2025},
  pages={356--371},
  year={2025}
}

@article{achiam2023gpt,
  title={Gpt-4 technical report},
  author={Achiam, Josh and Adler, Steven and Agarwal, Sandhini and Ahmad, Lama and Akkaya, Ilge and Aleman, Florencia Leoni and Almeida, Diogo and Altenschmidt, Janko and Altman, Sam and Anadkat, Shyamal and others},
  journal={arXiv preprint arXiv:2303.08774},
  year={2023}
}

@misc{anthropic_homepage,
  author       = {{Anthropic}},
  title        = {Home --- Anthropic},
  howpublished = {\url{https://www.anthropic.com/}},
  year         = {2025},
  note         = {Accessed: 2025-12-12}
}

@inproceedings{kamalloo2023evaluating,
  title={Evaluating open-domain question answering in the era of large language models},
  author={Kamalloo, Ehsan and Dziri, Nouha and Clarke, Charles and Rafiei, Davood},
  booktitle={Proceedings of the 61st annual meeting of the association for computational linguistics (volume 1: long papers)},
  pages={5591--5606},
  year={2023}
}

@inproceedings{zan2023large,
  title={Large language models meet nl2code: A survey},
  author={Zan, Daoguang and Chen, Bei and Zhang, Fengji and Lu, Dianjie and Wu, Bingchao and Guan, Bei and Yongji, Wang and Lou, Jian-Guang},
  booktitle={Proceedings of the 61st Annual Meeting of the Association for Computational Linguistics (Volume 1: Long Papers)},
  pages={7443--7464},
  year={2023}
}

@article{zhu2025large,
  title={Large language models for information retrieval: A survey},
  author={Zhu, Yutao and Yuan, Huaying and Wang, Shuting and Liu, Jiongnan and Liu, Wenhan and Deng, Chenlong and Chen, Haonan and Liu, Zheng and Dou, Zhicheng and Wen, Ji-Rong},
  journal={ACM Transactions on Information Systems},
  volume={44},
  number={1},
  pages={1--54},
  year={2025},
  publisher={ACM New York, NY}
}

@article{schick2023toolformer,
  title={Toolformer: Language models can teach themselves to use tools},
  author={Schick, Timo and Dwivedi-Yu, Jane and Dess{\`\i}, Roberto and Raileanu, Roberta and Lomeli, Maria and Hambro, Eric and Zettlemoyer, Luke and Cancedda, Nicola and Scialom, Thomas},
  journal={Advances in Neural Information Processing Systems},
  volume={36},
  pages={68539--68551},
  year={2023}
}

@article{wei2022chain,
  title={Chain-of-thought prompting elicits reasoning in large language models},
  author={Wei, Jason and Wang, Xuezhi and Schuurmans, Dale and Bosma, Maarten and Xia, Fei and Chi, Ed and Le, Quoc V and Zhou, Denny and others},
  journal={Advances in neural information processing systems},
  volume={35},
  pages={24824--24837},
  year={2022}
}

@article{yao2023tree,
  title={Tree of thoughts: Deliberate problem solving with large language models},
  author={Yao, Shunyu and Yu, Dian and Zhao, Jeffrey and Shafran, Izhak and Griffiths, Tom and Cao, Yuan and Narasimhan, Karthik},
  journal={Advances in neural information processing systems},
  volume={36},
  pages={11809--11822},
  year={2023}
}

@article{wang2024survey,
  title={A survey on large language model based autonomous agents},
  author={Wang, Lei and Ma, Chen and Feng, Xueyang and Zhang, Zeyu and Yang, Hao and Zhang, Jingsen and Chen, Zhiyuan and Tang, Jiakai and Chen, Xu and Lin, Yankai and others},
  journal={Frontiers of Computer Science},
  volume={18},
  number={6},
  pages={186345},
  year={2024},
  publisher={Springer}
}

@article{zhang2025survey,
  title={A survey on the memory mechanism of large language model-based agents},
  author={Zhang, Zeyu and Dai, Quanyu and Bo, Xiaohe and Ma, Chen and Li, Rui and Chen, Xu and Zhu, Jieming and Dong, Zhenhua and Wen, Ji-Rong},
  journal={ACM Transactions on Information Systems},
  volume={43},
  number={6},
  pages={1--47},
  year={2025},
  publisher={ACM New York, NY}
}

@online{gu2025hiringassistant,
  title        = {Building the agentic future of recruiting: how we engineered LinkedIn’s Hiring Assistant},
  author       = {Gu, Xiaoyang and Lu, Xie and Hewlett, Daniel},
  year         = {2025},
  month        = oct,
  day          = {21},
  organization = {LinkedIn Engineering},
  url          = {https://www.linkedin.com/blog/engineering/ai/how-we-engineered-linkedins-hiring-assistant},
  urldate      = {2025-12-12}
}

@article{liu2026simplemem,
  title={SimpleMem: Efficient Lifelong Memory for LLM Agents},
  author={Liu, Jiaqi and Su, Yaofeng and Xia, Peng and Han, Siwei and Zheng, Zeyu and Xie, Cihang and Ding, Mingyu and Yao, Huaxiu},
  journal={arXiv preprint arXiv:2601.02553},
  year={2026}
}

@article{lewis2020retrieval,
  title={Retrieval-augmented generation for knowledge-intensive nlp tasks},
  author={Lewis, Patrick and Perez, Ethan and Piktus, Aleksandra and Petroni, Fabio and Karpukhin, Vladimir and Goyal, Naman and K{\"u}ttler, Heinrich and Lewis, Mike and Yih, Wen-tau and Rockt{\"a}schel, Tim and others},
  journal={Advances in neural information processing systems},
  volume={33},
  pages={9459--9474},
  year={2020}
}

@article{comanici2025gemini,
  title={Gemini 2.5: Pushing the frontier with advanced reasoning, multimodality, long context, and next generation agentic capabilities},
  author={Comanici, Gheorghe and Bieber, Eric and Schaekermann, Mike and Pasupat, Ice and Sachdeva, Noveen and Dhillon, Inderjit and Blistein, Marcel and Ram, Ori and Zhang, Dan and Rosen, Evan and others},
  journal={arXiv preprint arXiv:2507.06261},
  year={2025}
}

@misc{gdpreu_guide,
  title        = {Complete guide to GDPR compliance},
  author       = {{GDPR.eu}},
  howpublished = {\url{https://gdpr.eu/}},
  note         = {Accessed: 2026-01-30}
}

@article{zhao20252graphrag,
  title={E\^{} 2GraphRAG: Streamlining Graph-based RAG for High Efficiency and Effectiveness},
  author={Zhao, Yibo and Zhu, Jiapeng and Guo, Ye and He, Kangkang and Li, Xiang},
  journal={arXiv preprint arXiv:2505.24226},
  year={2025}
}

@inproceedings{cheng2025ragtrace,
  title={Ragtrace: Understanding and refining retrieval-generation dynamics in retrieval-augmented generation},
  author={Cheng, Sizhe and Li, Jiaping and Wang, Huanchen and Ma, Yuxin},
  booktitle={Proceedings of the 38th Annual ACM Symposium on User Interface Software and Technology},
  pages={1--20},
  year={2025}
}

@inproceedings{maynez2020faithfulness,
  title={On faithfulness and factuality in abstractive summarization},
  author={Maynez, Joshua and Narayan, Shashi and Bohnet, Bernd and McDonald, Ryan},
  booktitle={Proceedings of the 58th Annual Meeting of the Association for Computational Linguistics},
  pages={1906--1919},
  year={2020},
  doi={10.18653/v1/2020.acl-main.173}
}

@inproceedings{zeng2024good,
  title={The good and the bad: Exploring privacy issues in retrieval-augmented generation (rag)},
  author={Zeng, Shenglai and Zhang, Jiankun and He, Pengfei and Liu, Yiding and Xing, Yue and Xu, Han and Ren, Jie and Chang, Yi and Wang, Shuaiqiang and Yin, Dawei and others},
  booktitle={Findings of the Association for Computational Linguistics: ACL 2024},
  pages={4505--4524},
  year={2024}
}

@article{jiang2024rag,
  title={Rag-thief: Scalable extraction of private data from retrieval-augmented generation applications with agent-based attacks},
  author={Jiang, Changyue and Pan, Xudong and Hong, Geng and Bao, Chenfu and Yang, Min},
  journal={arXiv preprint arXiv:2411.14110},
  year={2024}
}

@inproceedings{wu2024longmemeval,
  title={LongMemEval: Benchmarking chat assistants on long-term interactive memory},
  author={Wu, Di and Wang, Hongwei and Yu, Wenhao and Zhang, Yuwei and Chang, Kai-Wei and Yu, Dong},
  booktitle={The Thirteenth International Conference on Learning Representations},
  year={2025}
}

@article{li2025memos,
  title={Memos: A memory os for ai system},
  author={Li, Zhiyu and Xi, Chenyang and Li, Chunyu and Chen, Ding and Chen, Boyu and Song, Shichao and Niu, Simin and Wang, Hanyu and Yang, Jiawei and Tang, Chen and others},
  journal={arXiv preprint arXiv:2507.03724},
  year={2025}
}

\appendix

\section{\hltm Indexing-time Prompts}
\label{appendix:treemem_indexing_prompts}
\subsection{Facet Extraction}
\label{appendix_prompt_facet_extraction_offline}
\begin{lstlisting}[
    breaklines=true,
    basicstyle=\ttfamily\scriptsize,
    frame=single, framerule=0.5pt
]
<system message>
    You are a helpful assistant specializing in extracting structured information from the input text data. Only retain key information; no need to include all details (e.g., detailed activities).
    Return a JSON object with key \"facets\" and value as a flattened dictionary, in which the key is the facet name and the value is the corresponding extracted facet value in string format; no nested information.
    {{
     "facets": {{
            <facet_name>: <facet_value>,
            <facet_name>: <facet_value>,
            <facet_name>: <facet_value>,
            ...
        }}
    }}
</system message>

<user message> {document} </user message>
\end{lstlisting}

\label{ref_prompt_for_prompt_segmentation}

\newpage
\subsection{Question--Answer Generation}
\label{appendix_prompt_question_answer_generation_offline}
\begin{lstlisting}[
    breaklines=true,
    basicstyle=\ttfamily\scriptsize,
    frame=single, framerule=0.5pt
]
<system message>
    Analyze text, produce 5-10 self-contained Q&A pairs covering only the most critical info, and include a rationale explaining the result.
    Rules
    - Q&A must be explicit (no pronouns/ambiguous refs), concise, and essential-only.
    Output (single JSON)
    {{
      "rationale": "...",
      "question_answers": [
        {{ "question": "...", "answer": "...", "source": "<doc_id>" }},
        ...
      ]
    }}
</system message>

<user message>
    {document}
</user message>
\end{lstlisting}

\subsection{Summary Generation}
\subsubsection{Detailed Summary Generation}
\label{appendix_prompt_detailed_summary_generation}
\leavevmode\par
\begin{lstlisting}[
    breaklines=true,
    basicstyle=\ttfamily\scriptsize,
    frame=single, framerule=0.5pt
]
<system message>
    You are a helpful assistant. Summarize this input document to retain key information.
    Please include the document ID(s) in the output when provided.
</system message>

<user message>
    {document}
</user message>
\end{lstlisting}

\subsubsection{Concise Summary Generation}
\label{appendix_prompt_single_sentence_summary_generation}
\leavevmode\par
\begin{lstlisting}[
    breaklines=true,
    basicstyle=\ttfamily\scriptsize,
    frame=single, framerule=0.5pt
]
<system message>
    You are a helpful assistant. You are provided with the following data, please generate a single-sentence summary of this data, only retain key information.
</system message>

<user message>
    {detailed_summary}
</user message>
\end{lstlisting}

\subsection{Question Answering}
\label{appendix_prompt_question_answering}
\begin{lstlisting}[
    breaklines=true,
    basicstyle=\ttfamily\scriptsize,
    frame=single, framerule=0.5pt
]
<system message>
    You are a helpful assistant that generates an answer from a provided context.
    Output Format: a JSON object of this schema:
    {{
      "rationale": str, # rationale of generating the answer,
      "answer": str, # the answer to the query
      "citation": list[str], # a ranked list of node IDs provided in the context.
    }}
</system message>

<user message>
    Here is the query to be answered: 
    <query>
        {query}
    </query>

    Below is the context for answering the query:
    <context>
        {context}
    </context>
</user message>
\end{lstlisting}

\onecolumn

\section{Detailed Experiment Results }
\subsection{Performance Metrics}
\label{appendix:detailed_experiment_results}

Table~\ref{tab_appendix:performance_metrics} reports the detailed performance of \hltm and all baselines across five backbone models. \hltm achieves the best correctness on summary-style queries and the best F1 on retrieval-style queries in every model column, with most gains over the strongest baseline statistically significant under a Bonferroni-corrected Wilcoxon test. This indicates that \hltm provides robust, model-agnostic quality improvements rather than relying on a specific LLM backbone.

\begin{table}[h]
\centering
\small
\setlength{\tabcolsep}{4pt}
\caption{Performance metrics across models. Bold: best method per column. Underline: best baseline method per column. Stars: Wilcoxon signed-rank test of \hltm{} vs.\ best baseline, Bonferroni-corrected ($^{*}p{<}0.05$, $^{**}p{<}0.01$, $^{***}p{<}0.001$). $\uparrow$ indicates higher is better. SimpleMem and GraphRAG omitted due to excessive latency (see Table~\ref{tab_appendix:deployment_metrics}).}
\label{tab_appendix:performance_metrics}

\begin{tabular}{l ccccc | ccccc}
\toprule
& \multicolumn{5}{c|}{\textbf{Summary-style (Correctness $\uparrow$)}}
& \multicolumn{5}{c}{\textbf{Retrieval-style (F1 $\uparrow$)}} \\
\cmidrule(lr){2-6} \cmidrule(lr){7-11}
\textbf{Method}
& \textbf{GPT-4o} & \textbf{Gemma 3} & \textbf{Qwen3} & \textbf{GLM-4} & \textbf{GLM-4.5} & \textbf{GPT-4o} & \textbf{Gemma 3} & \textbf{Qwen3} & \textbf{GLM-4} & \textbf{GLM-4.5} \\
& \textbf{mini} & \textbf{27B} & \textbf{32B} & \textbf{32B} & \textbf{Air (106B)} & \textbf{mini} & \textbf{27B} & \textbf{32B} & \textbf{32B} & \textbf{Air (106B)} \\
\midrule
\textsc{Full-context}
  & $0.791$ & $0.629$ & $0.677$ & $0.638$ & $0.735$
  & $0.544$ & $0.440$ & $0.451$ & $0.422$ & $0.482$ \\
\textsc{RAG}
  & $0.770$ & $0.717$ & $0.738$ & $0.738$ & $0.759$
  & $0.604$ & $\underline{0.648}$ & $\underline{0.617}$ & $\underline{0.581}$ & $0.646$ \\
\textsc{Schema Filter}
  & $0.513$ & $0.741$ & $0.723$ & $0.744$ & $0.791$
  & $0.546$ & $0.562$ & $0.484$ & $0.412$ & $0.558$ \\
\midrule
\textsc{A-Mem}
  & $0.779$ & $0.739$ & $0.789$ & $0.774$ & $0.779$
  & $0.549$ & $0.514$ & $0.517$ & $0.475$ & $0.581$ \\
\textsc{HippoRAG}
  & $\underline{0.833}$ & $\underline{0.801}$ & $\underline{0.819}$ & $\underline{0.830}$ & $\underline{0.841}$
  & $0.560$ & $0.589$ & $0.549$ & $0.480$ & $0.535$ \\
\textsc{RAPTOR}
  & $0.715$ & $0.717$ & $0.712$ & $0.730$ & $0.722$
  & $0.348$ & $0.505$ & $0.535$ & $0.408$ & $0.516$ \\
\textsc{Mem0}
  & $0.521$ & $0.475$ & $0.613$ & $0.473$ & $0.501$
  & $0.369$ & $0.368$ & $0.336$ & $0.285$ & $0.366$ \\
\textsc{ReadAgent}
  & $0.796$ & $0.744$ & $0.762$ & $0.785$ & $0.813$
  & $\underline{0.617}$ & $0.575$ & $0.518$ & $0.485$ & $\underline{0.730}$ \\
\midrule
\rowcolor{gray!15}
\hltm (ours)
  & $\mathbf{0.892}^{***}$ & $\mathbf{0.850}^{***}$ & $\mathbf{0.874}^{***}$ & $\mathbf{0.867}^{*}$ & $\mathbf{0.893}^{***}$
  & $\mathbf{0.782}^{***}$ & $\mathbf{0.722}^{***}$ & $\mathbf{0.751}^{***}$ & $\mathbf{0.674}^{***}$ & $\mathbf{0.833}^{***}$ \\
\bottomrule
\end{tabular}

\end{table}

\subsection{Deployment Metrics}
\label{appendix_subsection_deployment_metrics}

Table~\ref{tab_appendix:deployment_metrics} compares \hltm's deployment efficiency against baselines on online query latency, LLM invocation count, and token consumption. \hltm keeps low query latency on both query types with a moderate token budget, which is substantially below full-context prompting, and lower or competitive against graph-based and other memory systems (GraphRAG, SimpleMem, HippoRAG, ReadAgent) despite using two LLM calls per query. This combination of high answer quality and bounded cost makes \hltm practical for latency-sensitive production deployment.

\begin{table}[h]
\centering
\small
\setlength{\tabcolsep}{5pt}
\caption{Deployment efficiency metrics across query types. Left: summary-style; right: retrieval-style. Metrics include latency, LLM calls, and token consumption. Mean $\pm$ SEM over queries. $\downarrow$ indicates lower is better.}
\label{tab_appendix:deployment_metrics}

\begin{tabular}{l c c c | c c c}
\toprule
& \multicolumn{3}{c|}{\textbf{Summary-style}}
& \multicolumn{3}{c}{\textbf{Retrieval-style}} \\
\cmidrule(lr){2-4} \cmidrule(lr){5-7}
\textbf{Method}
& \textbf{Latency (s) $\downarrow$}
& \textbf{LLM Calls $\downarrow$}
& \textbf{Tokens (k) $\downarrow$}
& \textbf{Latency (s) $\downarrow$}
& \textbf{LLM Calls $\downarrow$}
& \textbf{Tokens (k) $\downarrow$} \\
\midrule
\textsc{Full-context}
& $7.94 \pm 0.22$ & $1.00 \pm 0.00$ & $40.79 \pm 1.17$
& $10.23 \pm 0.35$ & $1.00 \pm 0.00$ & $37.95 \pm 1.61$ \\
\textsc{RAG}
& $3.30 \pm 0.13$ & $1.00 \pm 0.00$ & $2.53 \pm 0.02$
& $3.55 \pm 0.14$ & $1.00 \pm 0.00$ & $2.70 \pm 0.03$ \\
\textsc{Schema Filter}
& $3.55 \pm 0.15$ & $1.58 \pm 0.02$ & $7.62 \pm 0.43$
& $3.55 \pm 0.15$ & $1.67 \pm 0.02$ & $7.29 \pm 0.51$ \\
\midrule
\textsc{A-Mem}
& $4.58 \pm 0.13$ & $1.00 \pm 0.00$ & $7.21 \pm 0.17$
& $5.30 \pm 0.17$ & $1.00 \pm 0.00$ & $6.67 \pm 0.18$ \\
\textsc{HippoRAG}
& $6.87 \pm 0.14$ & $1.95 \pm 0.01$ & $11.40 \pm 0.18$
& $10.03 \pm 0.29$ & $1.94 \pm 0.02$ & $10.97 \pm 0.25$ \\
\textsc{RAPTOR}
& $2.26 \pm 0.05$ & $1.00 \pm 0.00$ & $0.53 \pm 0.00$
& $2.46 \pm 0.06$ & $0.93 \pm 0.01$ & $0.63 \pm 0.01$ \\
\textsc{GraphRAG}
& $11.11 \pm 0.16$ & $15.86 \pm 0.26$ & $30.30 \pm 0.52$
& $7.98 \pm 0.25$ & $15.20 \pm 0.43$ & $23.11 \pm 0.69$ \\
\textsc{SimpleMem}
& $14.50 \pm 0.15$ & $7.01 \pm 0.06$ & $6.78 \pm 0.13$
& $17.80 \pm 0.19$ & $7.44 \pm 0.06$ & $8.76 \pm 0.17$ \\
\textsc{Mem0}
& $2.24 \pm 0.05$ & $1.00 \pm 0.00$ & $0.28 \pm 0.01$
& $1.99 \pm 0.04$ & $1.00 \pm 0.00$ & $0.30 \pm 0.00$ \\
\textsc{ReadAgent}
& $6.51 \pm 0.18$ & $2.00 \pm 0.00$ & $11.67 \pm 0.32$
& $7.24 \pm 0.28$ & $2.00 \pm 0.00$ & $13.40 \pm 0.49$ \\
\midrule
\rowcolor{gray!15}
\hltm (ours)
& $3.01 \pm 0.06$ & $2.00 \pm 0.00$ & $3.91 \pm 0.05$
& $3.46 \pm 0.07$ & $2.00 \pm 0.00$ & $4.26 \pm 0.07$ \\
\bottomrule
\end{tabular}

\end{table}

\twocolumn

\clearpage
\balance
\section{Privacy Isolation Evaluation}
\label{section:privacy_isolation_experiments}

To empirically validate \hltm's tenant-scoped memory architecture, we conducted leakage experiments on LinkedIn's Hiring Assistant dataset. We indexed each memory system with a cross-tenant dataset comprising hiring projects from multiple recruiter seats and contracts, then queried each system with natural-language tenant-scope constraints. We measure leakage on two dimensions: (i) \textbf{Query-wise Leakage Ratio}: the fraction of queries returning at least one out-of-scope hiring project, (ii) \textbf{Entity-wise Leakage Ratio}: the average fraction of returned entities that are out-of-scope. As shown in Table~\ref{tab:leakage_experiment}, baseline methods exhibit substantial leakage, with RAPTOR and GraphRAG reaching 90\%. \hltm achieves 0\% leakage on both metrics, owing to its hierarchical tenant-scoped indexing that structurally enforces isolation boundaries rather than relying on post-hoc filtering.

\begin{table}[h]
\centering
\caption{Privacy leakage rates across memory systems on a cross-tenant hiring dataset. $\downarrow$ indicates lower value is better. \hltm achieves perfect isolation on both metrics.}
\label{tab:leakage_experiment}
\begin{tabular}{l >{\centering\arraybackslash}p{2.5cm} >{\centering\arraybackslash}p{2.5cm}}
\toprule
\textbf{Method} 
  & \makecell{\textbf{Query-wise} \\ \textbf{Leakage} $\downarrow$} 
  & \makecell{\textbf{Entity-wise} \\ \textbf{Leakage} $\downarrow$} \\
\midrule
\textsc{RAG}             & 64.0\%  & 54.6\% \\
\textsc{Schema Filter}         & 30.0\% & 36.2\% \\
\cmidrule(lr){1-3}
\textsc{A-Mem}          & 58.0\% & 48.5\%\\
\textsc{HippoRAG}        & 52.0\%  & 42.7\% \\
\textsc{RAPTOR}          & 90.0\%  & 86.3\% \\
\textsc{GraphRAG}        & 92.0\%  & 98.5\% \\
\textsc{SimpleMem}            &  30.0\% & 69.8\% \\   
\textsc{Mem0}            &  78.0\% & 69.4\% \\   
\textsc{ReadAgent}       & 88.0\%  & 77.0\% \\
\cmidrule(lr){1-3}
\cellcolor{gray!15}\hltm (ours) & \cellcolor{gray!15}$\mathbf{0.0\%}$ & \cellcolor{gray!15}$\mathbf{0.0\%}$ \\
\bottomrule
\end{tabular}
\end{table}

\section{Incremental Indexing Evaluation}
\label{appendix:incremental_indexing}

The lossless nature of \hltm's incremental updates follows directly from its tree-structured design, which exploits the independence between a node and its subtree siblings — updating one branch never affects others. We evaluated this property on LinkedIn's complete hiring dataset by comparing two indexing strategies: (a)~\emph{full indexing}, which rebuilds the entire tree from scratch, and (b)~\emph{incremental indexing}, which removes stale leaf nodes, appends newly created ones, and refreshes updated ones before propagating changes up their ancestor paths.

We ran both strategies in parallel over two weeks on real production data, measuring alignment between the two resulting indices in terms of both memory content and tree topology. The outputs were found to be identical and fully overlapping (accounting for LLM non-determinism), directly confirming losslessness. Beyond correctness, incremental indexing reduces end-to-end latency from over 7 days to approximately 12 hours, enabling frequent re-indexing and ensuring production data freshness. Having been deployed in the production pipeline for more than six months, it achieves an \textbf{80--90\% reduction in LLM invocations and cost} with no measurable degradation in output structure or memory quality.

\begin{table}[H]
\centering
\caption{Incremental indexing evaluation on LinkedIn's production dataset.}
\label{tab:incremental_indexing}
\setlength{\tabcolsep}{4pt}
\renewcommand{\arraystretch}{1.1}
\begin{tabular}{l c c}
\toprule
\textbf{Metric} & \textbf{Full} & \textbf{Incremental} \\
\midrule
Indexing Latency               & $\sim$7 days   & $\sim$12 hrs \\
LLM call reduction    & ---            & 80--90\% \\
Topology overlap        & ---  & identical \\
Index quality         & ---            & no degradation \\
\bottomrule
\end{tabular}
\end{table}

\section{Notation Summary}
\begin{table}[H]
\centering
\small
\caption{Notation summary.}
\label{tab:notation}
\begin{tabular}{l p{0.55\linewidth}}
\toprule
\textbf{Symbol} & \textbf{Meaning} \\
\midrule
$\mathcal{T} = (V, E)$      & HLTM rooted tree index. \\
$V = \{v\}$                  & All nodes $v$ (business entities). \\
$E = \{(p, c)\}$             & Directed parent--child edges. \\
$\mathcal{T}_{v}$            & Subtree rooted at $v$ (privacy / access scope). \\
$D_v$                        & Raw documents attached to $v$. \\
$M_v = (F_v, Q_v, S_v)$     & Memory triple at node $v$; $M_v = \mathcal{M}(v)$. \\
$F_v$                        & Facet view: salient key--value pairs at $v$. \\
$Q_v$                        & QA view: question--answer pairs at $v$. \\
$S_v$                        & Summary view: paragraph + one-sentence summary at $v$. \\
$q$                          & User query text. \\
$F_q$                        & Facet set parsed from $q$. \\
$s_{\text{facet}},\, s_{\text{QA}},\, s_{\text{summary}}(q,v)$
                             & Per-view retrieval scores for node $v$. \\
$k_{\text{facet}},\, k_{\text{QA}},\, k_{\text{summary}}$
                             & Top-$k$ cutoffs for each view. \\
$F_k(q),\, Q_k(q),\, S_k(q)$& Top-$k$ memories from facet / QA / summary views. \\
$V^\star$                    & Leaf nodes updated since last indexing cycle. \\
$\mathrm{emb}(\cdot)$        & Text-to-vector embedding function. \\
$\mathrm{LLM}(q, \text{ctx})$& Answer generation conditioned on $q$ and retrieved memories. \\
\bottomrule
\end{tabular}
\end{table}

\end{document}